\documentstyle[preprint,aps]{revtex}
\begin{document}

\draft

\title{Models for the integer quantum Hall effect: 
the network model, the Dirac equation, 
and a tight-binding Hamiltonian.}

\author{C.-M. Ho and J.T. Chalker}

\address{Theoretical Physics, University of Oxford,
1, Keble Road, Oxford OX1 3NP, 
United Kindom}

\date{April 19, 1996}

\maketitle

\begin{abstract}

We consider models for the plateau transition in the 
integer quantum Hall effect. Starting from the network model, 
we construct a mapping to the Dirac Hamiltonian in two dimensions. 
In the general case, the Dirac Hamiltonian has randomness in the mass, 
the scalar potential and the vector potential. Separately, we show that 
the network model can also be associated with a nearest neighbour, 
tight-binding Hamiltonian.

\end{abstract}
 

\section{Introduction}

Anderson localisation is central to understanding of
the integer quantum Hall effect (IQHE) \cite{qh}. 
In particular, the plateau 
transitions, between different quantised values for the Hall 
conductance,  
reflect delocalisation transitions in 
each Landau level. Scaling ideas \cite{khmelnitskii} 
provide a framework for understanding 
these transitions, and are supported by the results of experiment 
\cite{expt} 
and of numerical simulation \cite{qh}. Progress towards an analytical 
theory of 
the critical point, however, remains limited. 

The simplest starting point for such a theory is to neglect 
electron-electron interactions and consider a single particle 
moving in a magnetic field with a disordered impurity potential.
In pioneering work, Pruisken and collaborators \cite{P&K} obtained from 
this
a field-theoretic description, in terms of a $\sigma$-model. 
More recently, in response to the difficulties of  
extracting quantitative results from the $\sigma$-model,
several alternative formulations have been
explored: Read \cite{read}, Lee \cite{lee-spin-chain} 
and Zirnbauer \cite{zirnbauer}
have investigated spin chains; Lee and Wang \cite{wang} have
considered the replica limit of Hubbard chains; and Ludwig and
collaborators \cite{Ludwig} have discussed the Dirac equation.

The correspondence  between Dirac fermions in two space dimensions, and non-relativistic
charged particles moving in a magnetic field, stems from the fact that 
time-reversal symmetry is broken both by a mass term in the 
two-dimensional Dirac equation \cite{Ludwig,berry}, and by a magnetic field 
in the Schr\"odinger
equation. Moreover, as emphasised by Ludwig {\it et al}, the Hall 
conductance
of Dirac fermions, with fixed Fermi energy, has a jump of $e^2/h$,
if the fermion mass is tuned through zero. 
The critical behaviour at this transition
depends on the symmetries of the Hamiltonian.
The Dirac equation with
only a random vector potential is particularly
amenable to analysis \cite{Ludwig,randomA}, since the
zero-energy eigenstates are known explicitly \cite{A&C}.
Critical properties are controlled by a line of fixed
points, and turn out to be different from those
expected at plateau transitions in the IQHE.
The line of fixed points, however, is unstable against additional
randomness, either in the mass or in the scalar potential,
and flow is conjectured \cite{Ludwig}
to be towards a generic quantum Hall
fixed point, describing the same critical behaviour as emerges
from the usual Schr\"odinger equation.

Confidence that Dirac fermions with suitable randomness do indeed 
have a critical point in the same universality class as the IQHE 
plateau transitions
is clearly strengthened if there exists an explicit mapping from 
a microscopic model for the IQHE to the Dirac Hamiltonian. Fisher 
and Fradkin \cite{F&F}, and subsequent authors \cite{Ludwig,ziegler}, have reached 
the 
Dirac equation starting from certain, rather specific, tight-binding models.
An alternative to the tight-binding model, as a description of the 
IQHE, is the
network model \cite{network}, studied extensively by numerical 
simulation \cite{Lee}.
Ludwig and collaborators \cite{Ludwig} have asserted that Dirac fermions with 
the various possible kinds of randomness each represent particular 
forms of network model.
These authors, however, did not set out a transformation from one 
model to the other. 
Separately, Lee \cite{lee-spin-chain} found such a transformation, in the
particular case of a  network model 
without random phases, obtaining Dirac fermions with randomness 
only in the mass.

The purpose of this paper is to describe a general mapping 
from the network model
to the Dirac Hamiltonian in two dimensions, which, in the unrestricted
case,
has randomness in the mass, the scalar potential and the vector 
potential.
Any approach to this problem must confront the fact that the network
model is defined using the language of scattering theory, and therefore,
at least in the first instance, contains information only about
behaviour at one energy. The Dirac Hamiltonian, by contrast, 
obviously fixes properties of an entire spectrum of eigenstates.
We begin from a unitary matrix defined \cite{klesse}
for the network model, which,
heuristically, can be thought of as a time-evolution operator.
We show, in a continuum limit, that it is the evolution operator for
a Dirac Hamiltonian. In this respect, our route is rather
different from that of Lee \cite{lee-spin-chain}, who obtains a
Hamiltonian by endowing the phases of the network model with
an energy dependence. We also differ in taking the
continuum limit isotropically, while Lee \cite{lee-spin-chain} does so
anisotropically.

Our mapping is described in section II. In section III 
we examine in detail how edge states of the network model are related
to boundary states of Dirac fermions. This is important, since it is
these states which are responsible for the quantised Hall conductance 
away from plateau transitions. 

Equivalence between the network model and the Dirac 
Hamiltonian necessarily requires a continuum limit. 
In section IV, we show that, independently of the
continuum limit, one can associate with the network
model a tight-binding
Hamiltonian, which contains only nearest-neighbour hopping.

\section{Mapping from the network model to the Dirac Hamiltonian}

In this section we construct an explicit mapping from the
network model \cite{network} to the Dirac Hamiltonian in two 
dimensions.
First, we recall the physical basis for the network model
and its definition. Consider non-relativistic, charged
particles moving in a smoothly
varying scalar potential in two dimensions, with a strong 
perpendicular
magnetic field. The potential is smooth if its correlation
length is much larger than the cyclotron radius, and the field 
is
strong if the cyclotron energy is larger than the amplitude of
potential fluctuations. Under these conditions, the kinetic energy 
of 
cyclotron motion about the guiding centre, and the potential energy 
associated
with the position of the guiding centre, are both separately
conserved. We focus on drift
of guiding centres along equipotential lines.
In the network model, portions of a given equipotential are 
represented by
directed `links', and the wavefuntion for the particle is 
caricatured by
complex 
current amplitudes, $Z$, defined at points on each link. On 
traversing
a link, a particle aquires an Aharonov-Bohm phase: if $Z_i$ 
and $Z_j$ are
amplitudes at opposite ends of the link $k$ (see Fig.\ 1a),
$Z_{j}=e^{i\phi_{k}}Z_{i}$.
Tunneling between two disjoint 
portions of equipotential can occur where they are separated
by less than a cyclotron radius, as happens near saddle-points
in the potential. It is incorporated into the model
at `nodes', where two incoming and two outgoing 
links meet. The amplitudes on the four links that meet at a
given node may be related by a transfer matrix or
by a scattering matrix. In a suitable gauge, each of these 
$2 \times 2$ matrices is real
and depends on a single parameter, which we denote by $\theta$
(for the transfer matrix) and $\beta$ (for the scattering matrix).
The parameter determines the relative probabilities for a
particle to turn to the left or to the right on arriving at the 
node.
It is a smooth function of the equipotential energy, measured 
relative to
the potential at the saddle-point \cite{fertig}.
Referring to Fig.\ (1b), one has
\begin{equation}
\left( \begin{array}{c}
Z_4\\
Z_3
\end{array} \right) =
\left( \begin{array}{cc}
\cosh \theta & \sinh \theta \\
\sinh \theta & \cosh \theta
\end{array} \right)
\left( \begin{array}{c}
Z_1 \\
Z_2
\end{array} \right)
\label{transfer}
\end{equation}
and
\begin{equation}
\left( \begin{array}{l}
Z_2\\
Z_4
\end{array} \right) =
\left( \begin{array}{rr}
\cos\beta & \sin\beta \\
-\sin\beta & \cos\beta
\end{array} \right)
\left( \begin{array}{l}
Z_1\\
Z_3
\end{array} \right).
\label{scattering}
\end{equation}
The two parameters are related by 
$ \sin {\beta}=-\tanh {\theta}$. On varying the equipotential energy
from far below that of the saddlepoint to far above, 
$\beta$
increases from $\beta =0$ to $\beta= \pi/2$; tunneling is a 
maximum at
$\beta_c = \pi / 4$. 

The network model as a whole is built by connecting these
two elements - links and nodes - to form a lattice. The 
simplest choice is
the square lattice, illustrated in Fig.\ 2. Randomness is 
introduced by
choosing each link phase, $\phi_k$, 
independently from a probability distribution. The model 
represents particle motion at an energy determined by 
the value of the node parameters. If all nodes
are identical, and if phases are uniformly distributed 
between $0$ and $2 \pi$,
the system is critical at $\beta = \beta_c$,
and in the localised phase otherwise.

We follow Klesse and Metzler \cite{klesse}, and associate a 
unitary matrix with the model.
Roughly speaking, this matrix is a time 
evolution operator.
Let the unit of time be the interval required for a guiding 
centre
to drift from the mid-point of one link, through a node, to 
the mid-point 
of the next link; ignore dispersion in this time interval, 
arising from 
variations in drift velocity or in lengths of links.
Let $Z({\bf r};L)$ be the amplitude for a particle to arrive 
at a point, ${\bf r}$,
after $L$ time-steps, starting from an initial wavefunction  
$Z({\bf r'};0)$.
Then
\begin{equation}
Z({\bf r};L+1)=\sum_{{\bf r'}} T_{{\bf r,r'}} Z({\bf r'};L) 
\label{eqn4},
\end{equation}
and $T$ is the required time evolution operator. Eigenfunctions
of $T$ with eigenvalue $1$ are stationary states of the network
model.

In Eq\ \ref{eqn4}, the element $T_{{\bf r,r'}}$
is non-zero only if there is a one-step path on the 
lattice from ${\bf r}$ to ${\bf r'}$: that is, a path
which follows the directions of the links and passes 
only one node.
The values of these non-zero elements are given \cite{klesse} 
by a product of
a phase factor from the link traversed, and a tunneling 
amplitude
from the node, with sign conventions indicated in Fig.\ 3.

To be definite, consider the system illustrated in Fig.\ 4. 
Plaquettes are labelled by the coordinates, $(x,y)$, of their 
centres. 
With our choice of lattice constant and of orientation for the 
axes, $(x,y)$
are a pair of integers, either both even or both odd.
We denote the four links, $i$, making up a plaquette by $i=1,2,3$ 
and $4$,
so that a point, ${\bf r}$, on the network is specified by the 
combination
$(x,y,i)$.
Initially, we take the tunneling parameter, $\beta$, to be the 
same at every node, and the four phases, $\phi_i$, to be the
same on every plaquette. In addition, it is convenient to measure
the phases relative to their value when half a flux quantum threads
each plaquette, by replacing $\phi_4$ with $\phi_4 + \pi$.

Because each plaquette has four links, the matrix 
$T$ has a $4 \times 4$  block structure, and because the 
mid-points 
of links form a bi-partite lattice,
each such block consists of two $2 \times 2$ blocks.
To exhibit this structure, we arrange the amplitudes $Z({\bf r};L) 
\equiv Z_i(x,y;L)$
in the order $(Z_{+},Z_{-})$, with 
$Z_{+}(x,y) = (Z_1(x,y),Z_3(x,y))$ and 
$Z_{-}(x,y) = (Z_2(x,y),Z_4(x,y))$, suppressing 
the time, $L$, for clarity. In this basis, the evolution 
operator is
\begin{equation}
T= \left( \begin{array}{cc}
0 & M\\
N & 0
\end{array} \right),
\label{matrix}
\end{equation}
where 
\begin{equation}
M=
\left( \begin{array}{cc}
S e^{i\phi_{1}} t_{-}^{x}t_{+}^{y} & Ce^{i\phi_{1}}\\
Ce^{i\phi_{3}} & -S e^{i\phi_{3}} t_{+}^{x}t_{-}^{y}
\end{array} \right)
\end{equation}
and
\begin{equation}
N=
\left( \begin{array}{cc}
Ce^{i\phi_{2}} & S e^{i\phi_{2}} t_{+}^{x}t_{+}^{y}  \\
Se^{i\phi_{4}} t_{-}^{x}t_{-}^{y}& -Ce^{i\phi_{4}}
\end{array} \right). 
\end{equation}
Here, we have introduced  the abbreviations  $C = \cos{\beta}$ 
and $S = \sin{\beta}$, and the translation operators, 
$t_{\pm}^x$ and $t_{\pm}^y$,
defined by their action,
$t_{\pm}^{x}
Z_i(x,y)=Z_i(x\pm 1, y)$ and 
$t_{\pm}^{y}Z_i(x,y)=Z_i(x, y\pm 1)$.
The first row of $M$, for example, expresses the fact, 
illustrated in Fig.\ 4, that 
$Z_1(x,y;L+1) =  S e^{i \phi_1}  Z_2(x-1,y+1;L) +   C e^{i \phi_1} Z_4(x,y;L)$.

In order to decouple $Z_{+}$ from $Z_{-}$, 
we consider the two-step evolution
operator,
\begin{equation}
W \equiv T^2 =
\left( \begin{array}{cc}
MN & 0 \\
0 & NM
\end{array} \right).
\label{w}
\end{equation}
We may then deal just with the upper-left block, 
$U \equiv MN$, in the matrix $W$, and the 
component-pair $Z_{+}$.
Since, at this stage, we are treating a system 
without disorder,
$U$ is diagonalised by Fourier transform.
We write its eigenvectors, ${\bf u}$, as ${\bf u}^{\top}(x,y) = (v,w) e^{i(q_{x}x+q_{y}y)}$ and find 
\begin{equation}
U\, {\bf u}
= e^{-iV}
\left( \begin{array}{cc}
\gamma & \alpha \\
-{\alpha}^{\ast} & {\gamma}^{\ast}
\end{array} \right) {\bf u} 
= e^{i(\chi-V)}\,{\bf u}
\label{u}
\end{equation}
where
\begin{equation}
\begin{array}{l}
V = -\frac{1}{2}\sum_{j=1}^{4}\phi_{j}\,, \\
\gamma=
2SC\,e^{i[\frac{1}{2}(\phi_{1}-\phi_{3})-q_{x}]}\,
\cos(\frac{1}{2}(\phi_{2}-\phi_{4})
+q_{y})\,, \\  \alpha=
e^{i[\frac{1}{2}(\phi_{1}-\phi_{3})+q_{y}]}\,
[S^{2}e^{i[\frac{1}{2}
(\phi_{2}-\phi_{4})+q_{y}]}-C^{2}
e^{-i[\frac{1}{2}(\phi_{2}-\phi_{4})+q_{y}]}]\,,
\end{array} 
\end{equation}
and $\gamma^{\ast}$, $\alpha^{\ast}$ are the corresponding 
complex conjugates.
The eigenvalues of $U$ are $e^{i(\chi - V)}$ with a phase, 
$\chi$, given by
\begin{equation}
\cos{\chi} \equiv \sin{2\beta}\cos(q_{x}-A_{x})\cos(q_{y}-A_{y}),
\label{eigenvalue}
\end{equation}
in which $A_{x}=(\phi_{1}-\phi_{3})/2$ and 
$A_{y}=(\phi_{4}-\phi_{2})/2$. Setting 
$\beta = \beta_c +m/2 \equiv \pi/4 + m/2$, 
and taking $-\pi \leq \chi < \pi$, the range of 
allowed values for $\chi$ has gaps around $\chi=0$ 
and $\chi=\pm \pi$ for $m \not= 0$: $\chi$ satisfies 
$-\pi + |m| \leq \chi \leq - |m|$ or $|m| \leq \chi \leq \pi - |m|$. 
This dispersion relation is illustrated 
in Fig.\ 5. Stationary states of the network model are characterised by
$\chi - V = 0$.

To extract from the unitary evolution operator a Dirac Hamiltonian, 
$H$, we write 
$U = e^{-i\tilde{H}}$ and work in the continuum limit, in 
which $\tilde{H}$ is small.
Thus we
expand around $(q_{x},q_{y})=(0,0)$, taking $m$ and
the link phases, $\phi_{j}$, to be small. To leading order, Eq.(\ref{eigenvalue}) gives the spectrum
\begin{equation}
\chi^{2}=m^{2}+(q_{x}-A_{x})^{2}+(q_{y}-A_{y})^{2}
\label{}
\end{equation} for small $\chi$. At an operator level, when 
$t_{\pm}^x$ and $t_{\pm}^y$ act on smooth functions we  make 
the replacements
$t_{\pm}^x = 1 \pm \partial_{x}$ and $t_{\pm}^y = 1 \pm \partial_{y}$.
Then 
\begin{equation}
U \approx 1 + \left( \begin{array}{lr}
-\partial_{x}+iA_{x} & \partial_{y}-i A_{y} + m \\
\partial_{y}-i A_{y} - m & \partial_{x}-i A_{x}
\end{array} \right) -iV \equiv 1 - i \tilde{H}
\label{continuum}
\end{equation}
Finally, to bring the Hamiltonian into a conventional form,
we make a rotation in the two-component space, setting
$H=G\tilde{H}G^{-1}$, with 
\begin{equation}
G =   \frac{1}{\sqrt 2}\left( \begin{array}{cc}
i& -1 \\
i & 1
\end{array} \right)
\end{equation}
and obtain
\begin{equation}
{H=( p_{x} - A_{x} )\sigma_{x}+( p_{y} - A_{y} )\sigma_{y} + 
m \sigma_{z} + V }\,,
\label{dirac1}
\end{equation} where  $p_{x}= -i \partial_{x}$, and similarly 
for $p_{y}$,
and we use the Pauli matrix representation 
\begin{equation}
\sigma_{x}=\left( \begin{array}{rr}
0 & 1\\
1 & 0
\end{array} \right)$ ,\ \ $ \sigma_{y}=\left( \begin{array}{cc}
0 & -i\\
i & 0
\end{array} \right)$ ,\ \ $ \sigma_{z}=\left( \begin{array}{rr}
1 & 0\\
0 & -1
\end{array} \right) \,.
\end{equation}

Now consider a network model with randomness. If the link phases 
and tunneling parameter vary smoothly in space, one obtains in 
the continuum limit the Dirac Hamiltonian, Eq.\ (\ref{dirac1}), 
with  randomness in the vector potential, scalar potential and 
mass. Specifically, fluctuations in the vector potential, 
${\bf A}$, arise from randomness in the individual link phases, 
fluctuations in the scalar potential, $V$, come from variations 
in the total Aharonov-Bohm phase
associated with each plaquette, and fluctuations in the mass, $m$, 
are present if the tunneling parameter is not constant everywhere.
The time-independent states of the network model correspond to
the zero-energy states of the Dirac Hamiltonian.

This mapping can also be carried through for generalisations of the
network model. In particular, the two-dimensional model in
which each link carries $N$ channels \cite{network} is equivalent
to the $U(N)$ Dirac Hamiltonians investigated by Fradkin \cite{fradkin}.

\section{Edge states in the network model and the Dirac equation}

The edge of a sample is, of course, set by a scalar confining 
potential in the usual description of the IQHE, based on the
Schr\"odinger equation.  Dirac fermions, by contrast, are confined
by a spatially dependent mass, as discussed by 
Ludwig {\it et al} \cite{Ludwig}.
In particular, chiral, zero-energy states of Dirac fermions
are associated with contours of zero mass \cite{Ludwig}. We discuss 
in this section
how such edge states emerge in our mapping 
from the network model to the Dirac Hamiltonian.

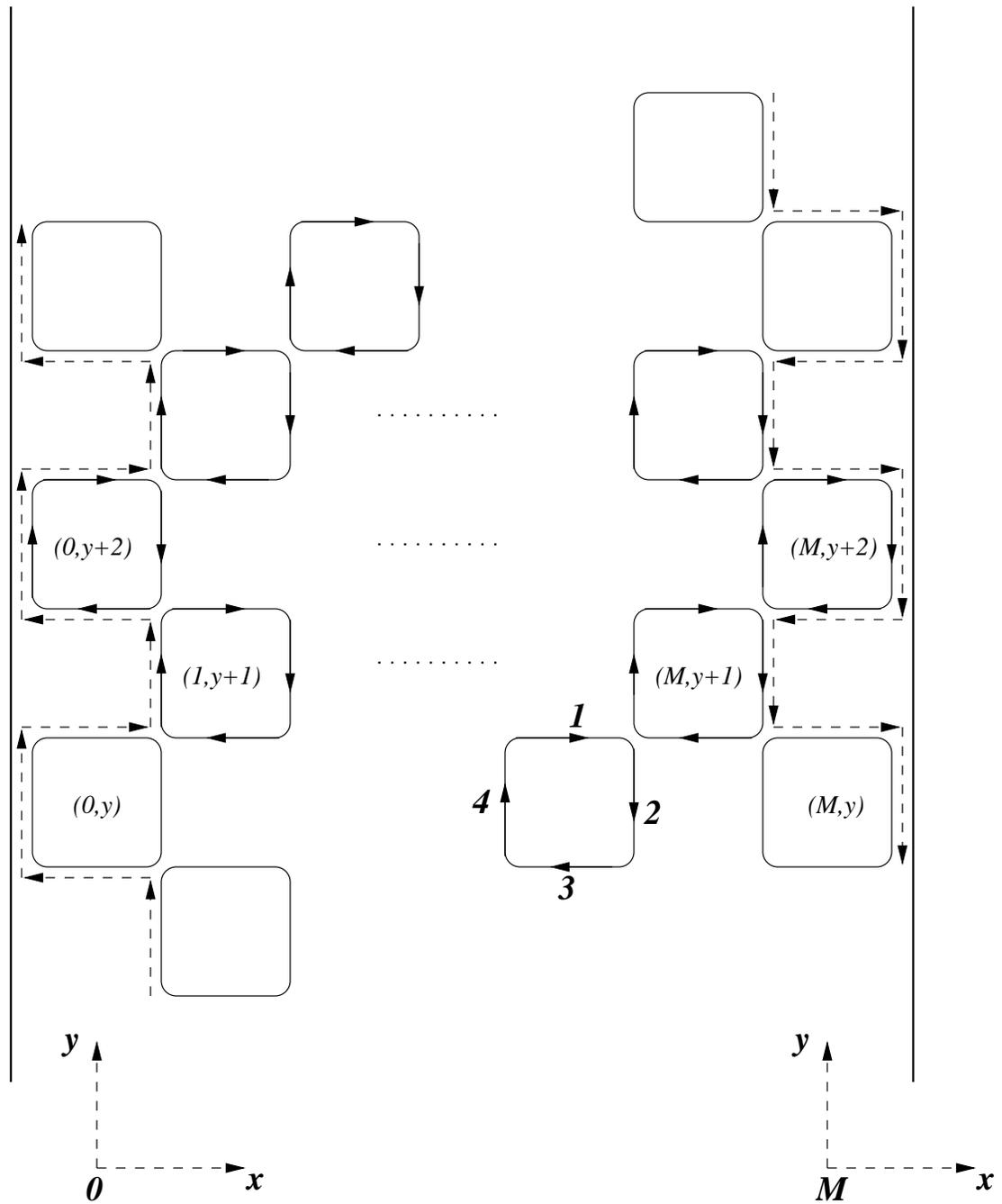
\begin{figure}
\caption{The network model defined on a strip of width $M$.
Dashed arrows indicate the propagation direction of edge states.
The dotted arrow represents the boundary condition, Eq\ 17.} 
\label{strip}
\end{figure}

Consider a network model defined on a strip of finite width, as in
Fig.\ \ref{strip}. For energies in the lower half of the Landau level,
corresponding to values of the node parameter $\beta < \beta_c$, all
states are localised, while for energies in the upper half of the
Landau level, for which $\beta > \beta_c$, a pair of extended states
appears, one at each edge of the strip \cite{network}.

The evolution operator, $U$, acting on the two-component wavefunction,
$Z_+(L)$, introduced for the bulk of the system in the preceeding
section, is supplemented at the edge by the following
boundary conditions (see Fig.\ \ref{strip}):
at $x=0$, the component $Z_3$ obeys the same equation
as in the bulk,
\begin{equation}
Z_3(0,y;L+2) = [U\, Z_+(L)]_3(0,y),
\label{bdy3}
\end{equation}
while the component $Z_1$ satisfies
\begin{equation}
Z_1(0,y;L+2)= - e^{i[\phi_1(0,y)+\phi_4(0,y)]}\, Z_3(0,y,L).
\label{bdy1}
\end{equation}
We wish to check under what conditions the evolution operator, $U$, 
has an eigenvector, ${\bf u}$ representing an edge state.
We simplify the discussion by considering a semi-infinite system 
without
disorder, setting $\phi_i = 0$ for all $i$. Without
disorder, the spatial dependence, for $x \ge 1$, of such an 
eigenvector is
${\bf u}^{\top}(x,y) = (v,w) e^{(iq y - \lambda x)}$,
where $Re [ \lambda] > 0$; for $x = 0$, 
${\bf u}^{\top}(0,y) = (v',w') e^{iq y}$, where $v,v',w$ and $w'$ are 
constants.
This ansatz in the equation
\begin{equation}
U \, {\bf u} = e^{i\chi} \, {\bf u}\, ,
\end{equation}
taking $U$ from Eqs.\ (\ref{w}),\ (\ref{bdy3}) and (\ref{bdy1}),
yields
\begin{equation}
e^{\lambda}= \tan \beta
\end{equation}
confirming that an edge sate exists ($Re [ \lambda ] > 0$)
only for $\beta > \beta_c \equiv \pi/ 4$.

Similar results also follow if one considers directly the continuum 
limit.
Let the eigenfunctions of $\tilde{H}$ be $\tilde{\Psi}(x,y)$,
so that those of $H$ are $\Psi(x,y)$, with 
$\Psi \equiv G \tilde{\Psi}$.
Writing $\Psi^{\top} = (\Psi_1, \Psi_3)$, the boundary
condition at $x=0$, Eq.\ (\ref{bdy1}), is to leading order
\begin{equation}
\Psi_3(0,y) =  i \Psi_1(0,y).
\label{boundry}
\end{equation}
Note that this boundary condition enforces a chiral edge current:
the current density, with
components $j_{\alpha} = \Psi^{\dagger} \sigma_{\alpha} \Psi$,
is ${\bf j} = ( 0, |\Psi_1|^2 )$, and necessarily in the 
positive $y$-direction.
Imposing this boundary condition, the Dirac Hamiltonian, $H$, 
of Eq.\ (\ref{dirac1}) with $V=0$ and ${\bf A} ={\bf 0}$
has an eigenfunction of energy $E$
\begin{equation}
\Psi(x,y)= 
e^{iEy}e^{-mx} {\left( \begin{array}{c}
1 \\
i
\end{array} \right)\ } 
\end{equation}
provided that $m > 0$.
This same eigenstate appears from considering an infinite 
system with
position-dependent mass, $m(x,y)$, following Ludwig {\it et al}: 
setting $m(x,y)=m$ for $x>0$, and $m(x,y) =m_0$ for $x<0$,
the boundary condition, Eq.\ (\ref{boundry}), emerges in the 
limit $m_0 \to - \infty$.

\section{Mapping from the network model to a tight-binding model}

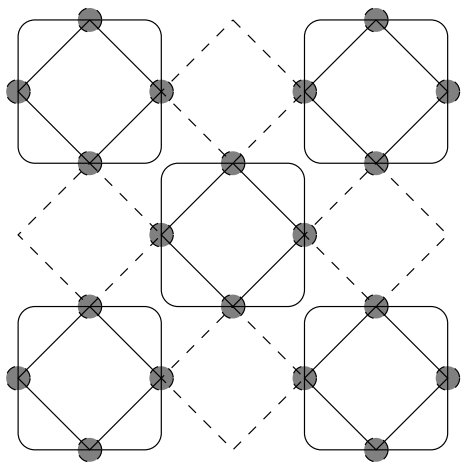
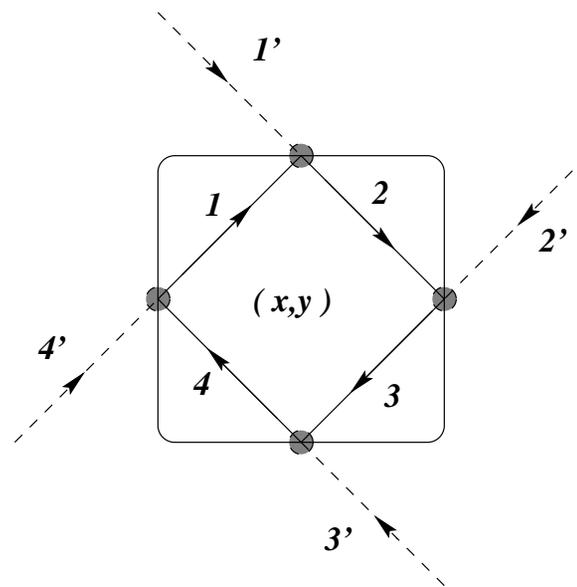
\begin{figure}
\caption{Schematic illustration of the tight-binding 
Hamiltonian. (a) The only non-zero matrix elements are 
those linking nearest-neighbour sites,
within plaquttes (full lines) and between plaquettes (dashed lines).
(b) Their values are: $ie^{i\phi_l(x.y)} \cos\beta$ 
for the bonds marked $l=1,2,3,4$;  $ie^{i\phi_l(x.y)} \sin\beta$,
for the bonds marked $l = 1',2'$; and $-ie^{i\phi_l(x.y)} \sin\beta$,
for the bonds marked $l=3',4'$, hopping in all cases being in the
direction given by the arrows.}
\label{tight-binding}
\end{figure}

It is also possible, without taking a continuum limit, to 
associate a nearest-neighbour tight-binding Hamiltonian with 
the network model.
The sites of the tight-binding model, each carrying one basis 
state, correspond to the links of the network model.
In terms of the one-step evolution matrix, $T$, the tight binding 
Hamiltonian, ${\cal H}$, is simply 
\begin{equation}
{\cal H} = (T^{\dag}-T)/i .
\end{equation} 
We indicate schematically in Fig.\ \ref{tight-binding} which
matrix elements of ${\cal H}$ are non-zero, and give their values 
in terms of link phases and the tunneling parameter.

This Hamiltonian has two important features. First, it is natural 
to introduce a unit cell containing the four sites arising from one
plaquette of the network model. The amplitudes of nearest neighbour 
hopping {\it within} and {\it between} unit cells have moduli  
$\cos\beta$ 
and $\sin\beta$, 
respectively, where $\beta$ is the tunneling parameter: they are 
different, except at the critical point, 
$\beta=\beta_c \equiv \pi/4$. Second, the phases of the hopping 
matrix elements are correlated in the way indicated in 
Fig.\ \ref{tight-binding}. 
It follows from known behaviour of the network model that these 
correlations have unusual consequences for the tight-binding 
model. Consider a system in which all link phases are 
independently and uniformly distributed. 
It is straightforward to see that for this system the eigenvalues 
of $T$ in the complex plane are distributed uniformly on the unit 
circle. As a result, the density, $\rho(E)$, of eigenvalues, $E$, 
of ${\cal H}$ can be given exactly: $\rho(E)=1/4\pi[1-(E/2)^2]^{1/2}$ 
for $(E/2)^2 \leq 1$ and $\rho(E)=0$ for $(E/2)^2 >1$.
In addition, one can see that the 
eigenvectors of $T$ have the same statistical properties throughout 
the spectrum, a feature inherited by ${\cal H}$. Hence
the localisation length of eigenstates 
of ${\cal H}$ is independent of their energy, $E$.
If all nodes have the same parameter value, $\beta$, then as 
$\beta \to \beta_c$, the localisation length diverges 
uniformly across the spectrum: this Hamiltonian never 
has a mobility edge as a function of energy.

We note that ${\cal H}$ is similar in structure to, 
but different in detail from the tight-binding model used 
as a depature point by Ludwig and 
collaborators \cite{Ludwig}.
The latter includes not only nearest-neighbour, 
but also next-nearest
neighbour hopping: compare Fig.\ \ref{tight-binding} 
with Fig.\ 1 of Ref\ \cite{Ludwig}. Our model also differs 
from that of Fisher and Fradkin \cite{F&F}. 

\section{Summary}

We have set out in detail a mapping, from the network model
for plateau transitions in the IQHE, to Dirac fermions in
two space dimensions. The mapping makes crucial use of 
a unitary operator defined for the network model \cite{klesse},
which is essentially the time-evolution operator.
The two-component structure of Dirac spinors
in two space dimensions arises rather naturally from the
network model, defined on a square lattice: 
the fact that each plaquette has four sides
suggests a four-component
wavefunction, which separates into two independent
pairs because of the existence of two sublattices.
This structure is not dependent on the
continuum limit, and is also shared by a nearest-neighbour
tight-binding Hamiltonian, derived directly
from the evolution operator.

\section*{acknowledgement}
This work was supported in part by EPSRC grant GR/Go 2727 
and by an ORS award to C.-M. H. from the CVCP.

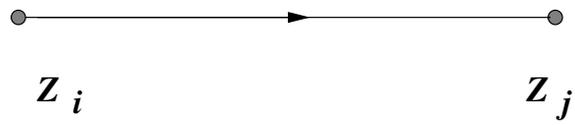
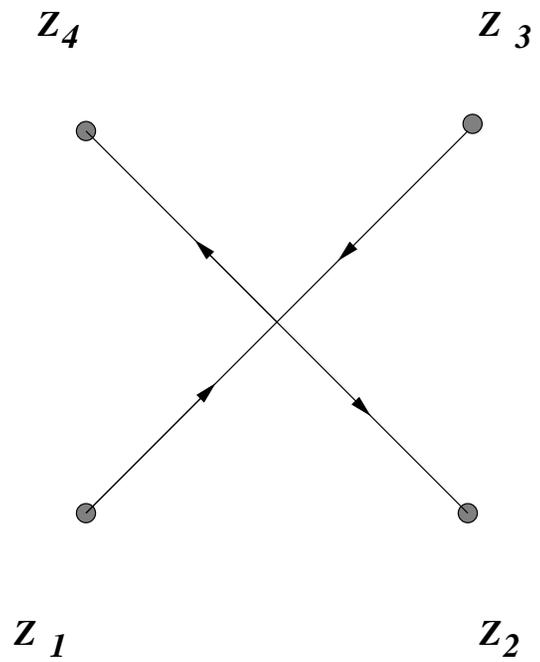
\begin{figure} 
\caption{ Components of the network model:
links (a), and nodes (b).}
\label{fig1}  
\end{figure}

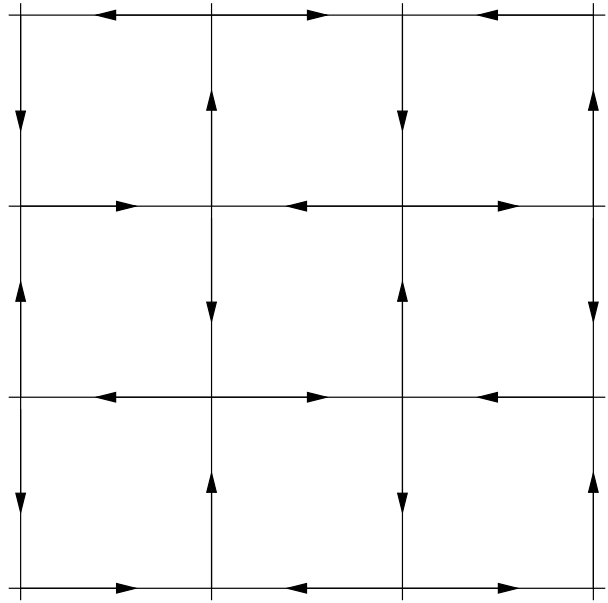
\begin{figure}
\caption{The network model on a square lattice.}
\end{figure}

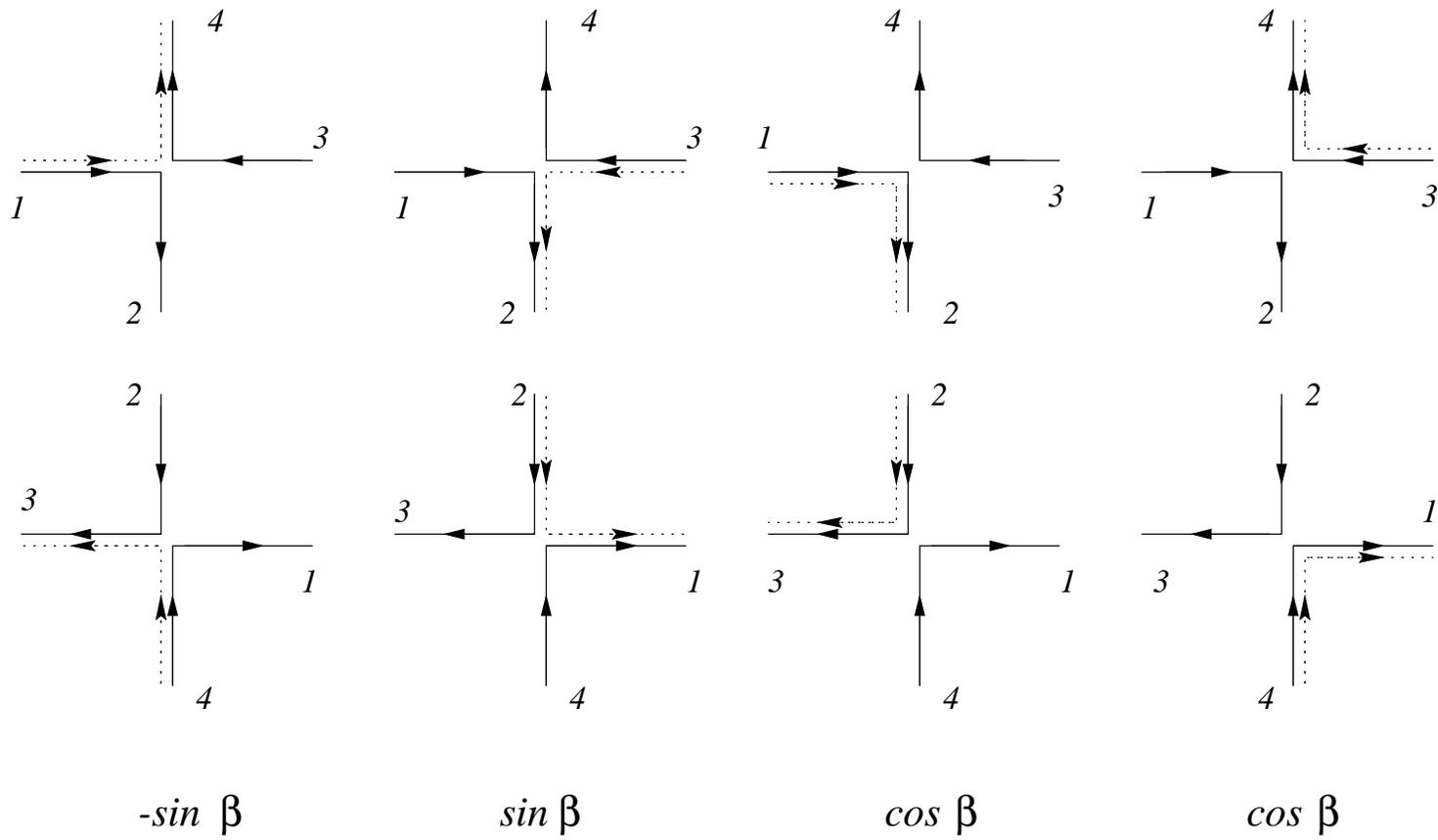
\begin{figure}
\caption{Amplitudes associated with possible scattering 
paths at nodes. }
\end{figure}

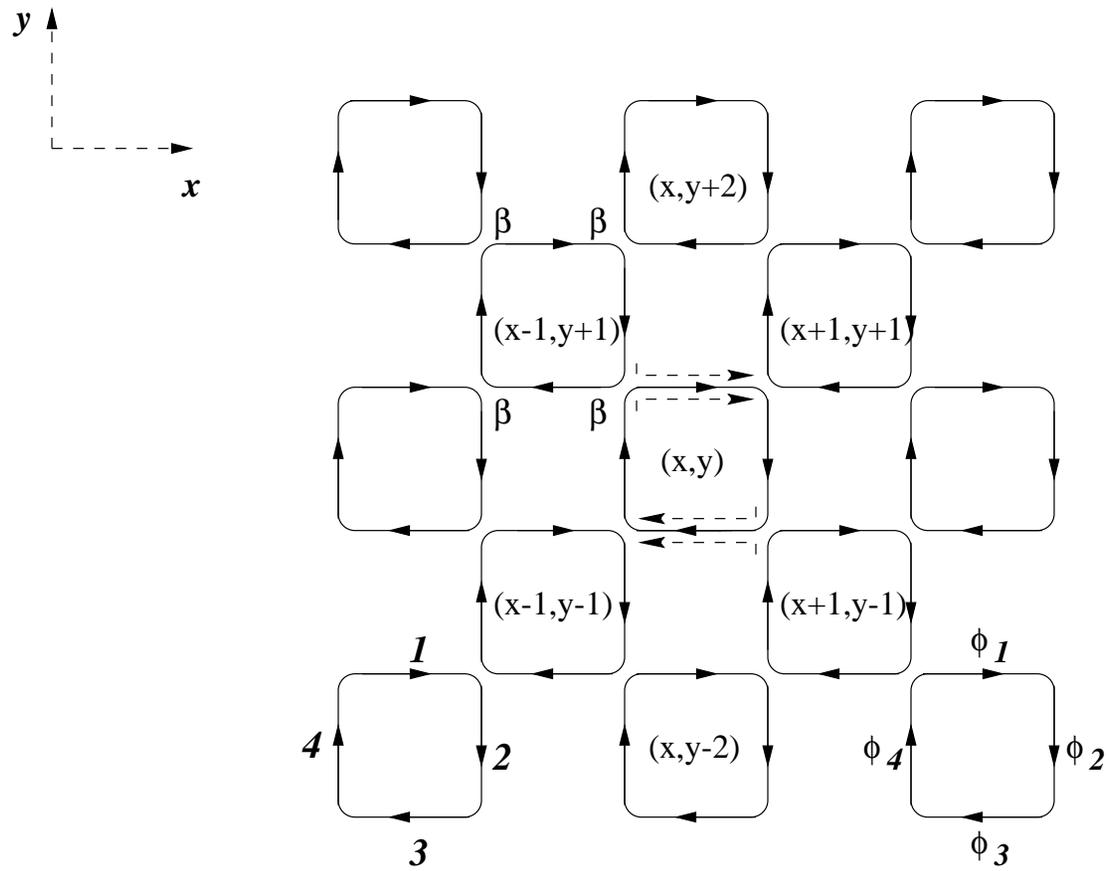
\begin{figure}
\caption{The network model, showing: (i) our coordinate system
for plaquttes; (ii) labelling of the four links that
make up a plaquette; and (iii), with dashed lines,
 the paths that contribute to
$Z_i(x,y;L+1)$ for $i=1,3$.}  
\end{figure}

\begin{figure}
\caption{The dispersion of $\chi$ ( vertical axis ) plotted as
a function of $(p-A)_x$ and $(p-A)_y$ (horizontal plane)
in units of $\pi$. The width of the gap between the two bands is 
determined by the mass which is 1.2 here.}      
\end{figure}

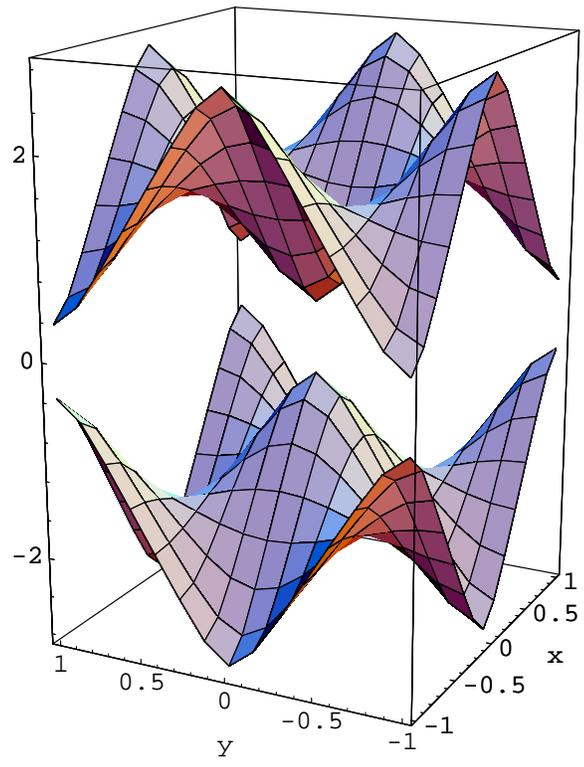
\end{document}